\documentclass[reprint,aps,prl,amsmath,amssymb,amsfonts,dvips]{revtex4-1}

\usepackage[dvips]{graphicx,color}
\usepackage{braket,bm}
\usepackage{dcolumn}
\begin{document}


\title{Randomness-induced quantum spin liquid behavior in the $s=$1/2 bond-random Heisenberg antiferromagnet on the pyrochlore lattice}


\author{Kazuki Uematsu}
\email[]{uematsu@spin.ess.sci.osaka-u.ac.jp}
\affiliation{Department of Earth and Space Science, Graduate School of Science, Osaka University, Toyonaka, Osaka 560-0043, Japan}
\author{Hikaru Kawamura}
\email[]{kawamura@ess.sci.osaka-u.ac.jp}

\affiliation{Department of Earth and Space Science, Graduate School of Science, Osaka University, Toyonaka, Osaka 560-0043, Japan}

\date{\today}

\begin{abstract}
 We investigate the zero- and finite-temperature properties of the bond-random $s=1/2$ Heisenberg antiferromagnet on the pyrochlore lattice by the exact diagonalization and the Hams--de Raedt methods. We find that the randomness induces the gapless quantum spin liquid (QSL) state, the random-singlet state. Implications to recent experiments on the mixed-anion pyrochlore-lattice antiferromagnet Lu$_2$Mo$_2$O$_5$N$_2$ exhibiting gapless QSL behaviors are discussed.
\end{abstract}


\maketitle

The quantum spin liquid (QSL) state has attracted much attention as an exotic state of matter, where any magnetic long-range order (LRO) or a spontaneous symmetry breaking is absent down to low temperatures due to strong quantum fluctuations. In the quest for the QSL state, geometrical frustration has been examined quite extensively as a key ingredient. In the last decade, a variety of QSL candidates were experimentally reported in geometrically frustrated quantum magnets in two dimensions (2D), {\it e.g.\/}, $s=1/2$ organic salts on the triangular lattice-like $\kappa$-(ET)$_2$Cu$_2$(CN)$_3$ \cite{ET-Kanoda, ET-Nakazawa,ET-Matsuda,ET-Jawad,ET-Sasaki}, EtMe$_3$Sb[Pd(dmit)$_2$]$_2$ \cite{dmit-Itou,dmit-Matsuda,dmit-Nakazawa,dmit-Jawad}, $\kappa$-H$_3$(Cat-EDT-TTF)$_2$ \cite{Isono1,Isono2}, and $s=1/2$ inorganic kagome antiferromagnet herbertsmithite CuZn$_3$(OH)$_6$Cl$_2$ \cite{Shores,Helton,Olariu,Freedman,Han,Imai}. Most of these QSL magnets are $s=1/2$ Heisenberg magnets exhibiting gapless (or nearly gapless) QSL behaviors with the $T$-linear low-temperature ($T$) specific heat. Gapless QSL behaviors have been observed not only for geometrically frustrated lattices, but also for geometrically unfrustrated lattices such as square and honeycomb lattices. In the latter, frustration is borne by, {\it e.g.\/}, the competition between the nearest- and  next-nearest-neighbor interactions, $J_1$ and $J_2$. Examples might be a square-lattice magnet Sr$_2$Cu(Te$_{1-x}$W$_{x}$)O$_6$ \cite{Mustonen,Mustonen2,Tanaka} and a honeycomb-lattice magnet 6HB-Ba$_3$NiSb$_2$O$_9$ \cite{Cheng,Mendels-Ni}. In this way, there are now considerable number of experimental realizations of gapless QSL in 2D. The true physical origin of such gapless QSL behaviors, however, still remains controversial.

 One promising scenario might be that the randomness or inhomogeneity, either of extrinsic or intrinsic origin, induces the gapless QSL-like state \cite{Watanabe,Kawamura,Shimokawa,Uematsu,Uematsu2}. The present authors and collaborators have demonstrated that such a randomness-induced gapless QSL-like state, called the ``random-singlet state,'' is stabilized for various 2D frustrated $s=1/2$ Heisenberg models, including the triangular \cite{Watanabe,Shimokawa}, kagome \cite{Kawamura,Shimokawa}, $J_1-J_2$ square \cite{Uematsu2} and $J_1-J_2$ honeycomb \cite{Uematsu} magnets, 
as long as the randomness is moderately strong. The origin of such (effective) randomness could be of variety, {\it e.g.\/}, the intrinsic ones like the dynamical freezing of the charge (dielectric) degrees of freedom in case of $\kappa$-ET and dmit salts and the slowing down of the proton motion in case of Cat salt, or the extrinsic ones like the possible Jahn-Teller distortion accompanied by the the random substitution of Zn$^{2+}$ by Cu$^{2+}$ in case of herbertsmithite and the random occupation of Te/W, in case of Sr$_2$Cu(Te$_{1-x}$W$_{x}$)O$_6$. The role of randomness on the possible QSL-like behavior was also studied recently by other groups for various 2D Heisenberg models \cite{Kimchi,Kimchi2,Sheng,Guo}. 

\begin{figure}[t]
  \includegraphics[width=\hsize]{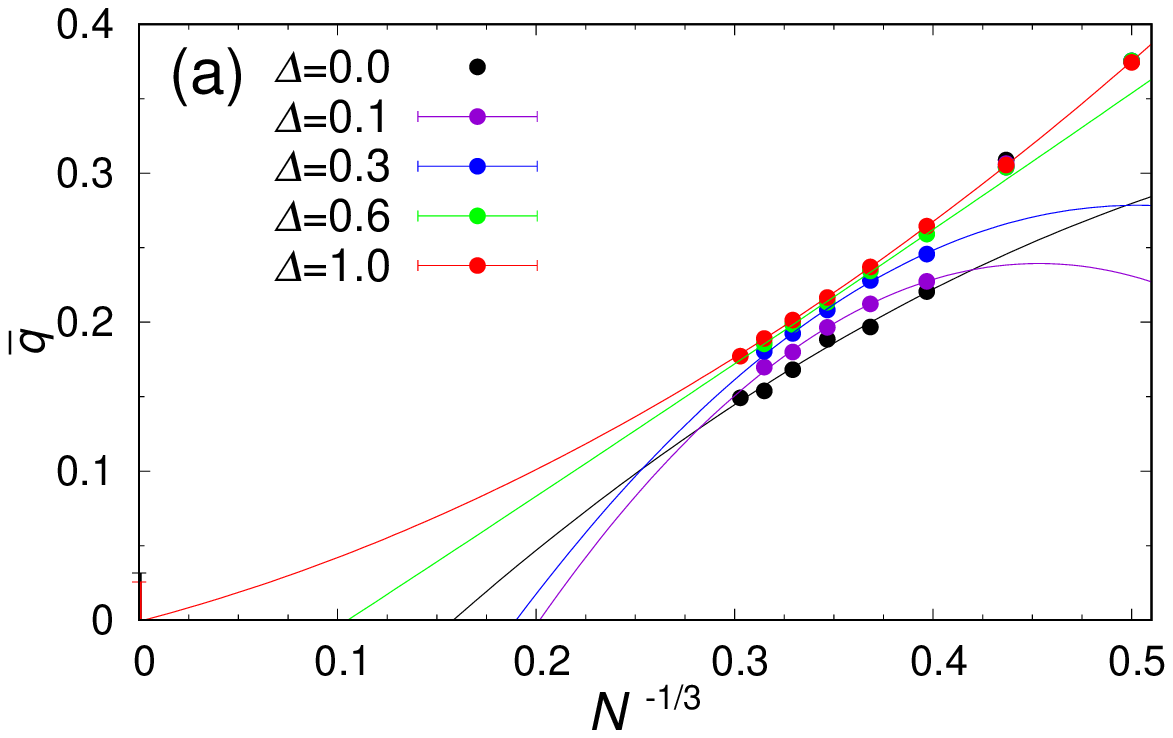}\\
  \includegraphics[width=\hsize]{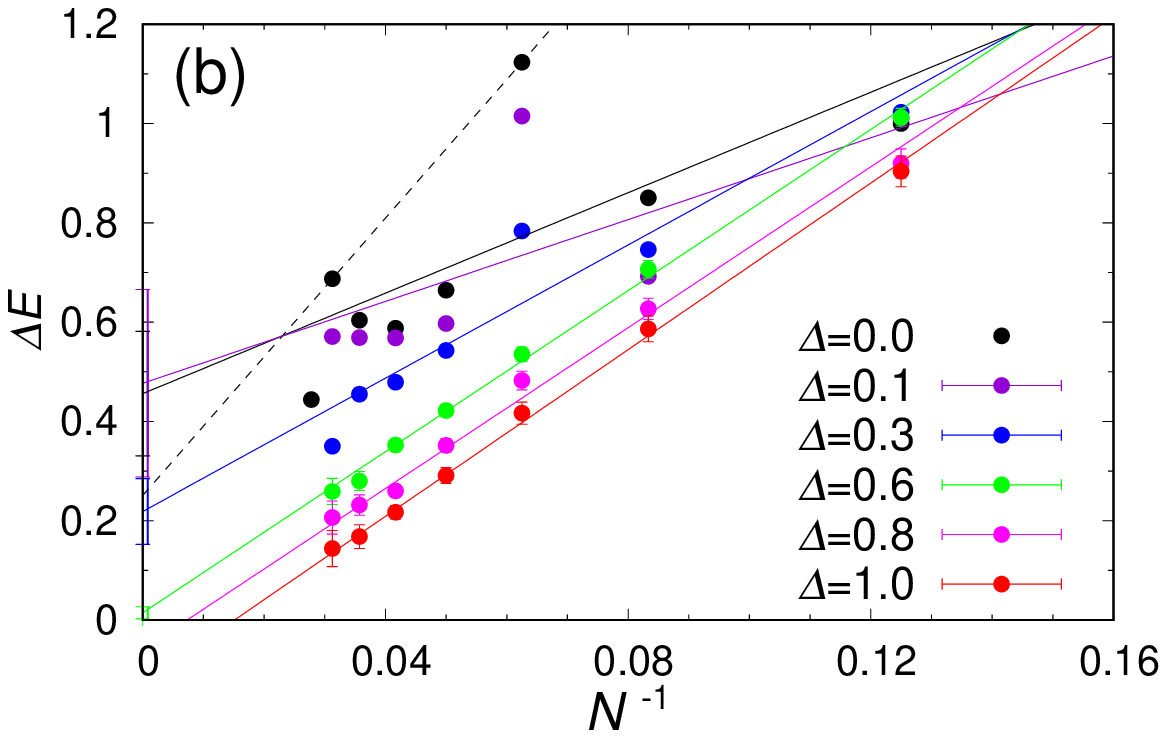}\\
  \includegraphics[width=\hsize]{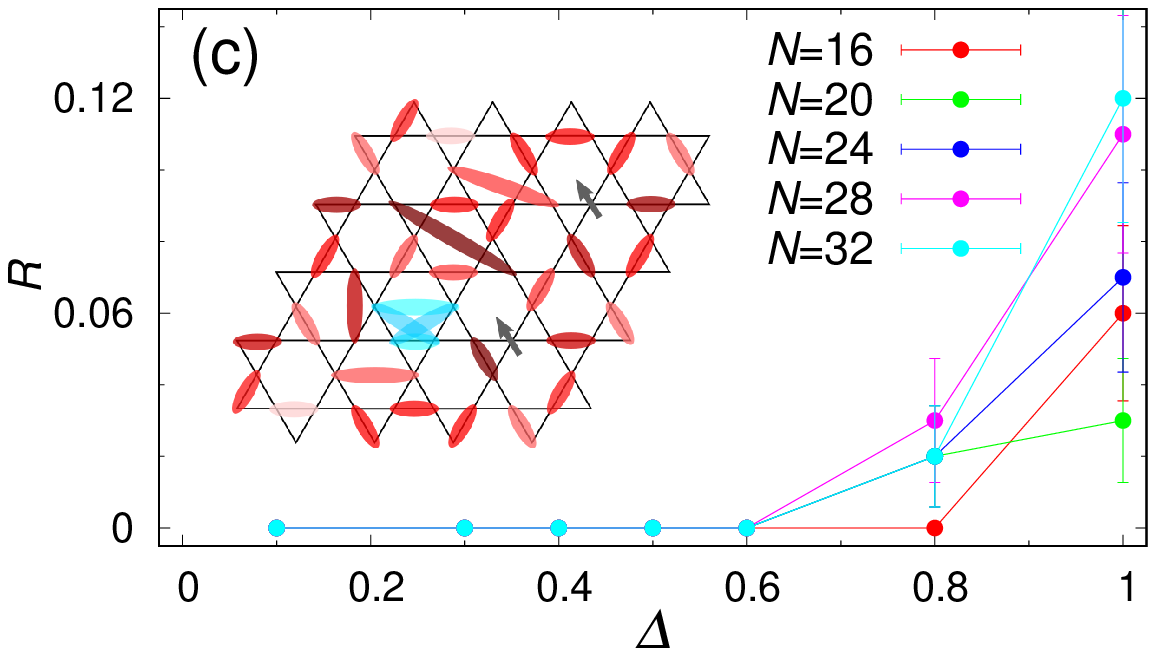}\\
  \caption{\label{fig:order} (Color online) (a) The spin freezing parameter $\bar{q}$ plotted versus $1/{N}^{1/3}$ for various values of $\Delta$. Lines are quadratic fits of the data for $N\geq16$. 
    (b) The mean spin-gap energy $\Delta E$ for various values of $\Delta$ plotted versus $1/N$. The solid lines are linear fits of all the data, while the dashed lines are linear fits of the data for the cubic-symmetric clusters, $N=16$ and $32$.
    (c) The ratio of the samples with triplet ground states plotted versus $\Delta$. Inset is the schematic picture of the random-singlet state. Details are described in the main text.} 
\end{figure}

The random-singlet state may roughly be pictured as the random covering of singlet-dimers to  minimize the total energy for a given random distribution of the exchange couplings $\{ J_{ij}\}$, as schematically illustrated in the inset of Fig. \ref{fig:order}(c). Such dimer covering on the random lattice, however, is highly nontrivial, leading to not only nearest-neighbor (NN) singlets but also further-neighbor singlets (represented by red dimers with their strength denoted by their thickness), to nearly-free spins failed to form singlets, the so-called ``orphan spins'' (represented by arrows), and even to resonances between distinct singlet-dimers (represented by blue dimers).  Low-energy excitations of the state would be (i) local singlet-to-triplet excitations of weakly-bound singlets, (ii) recombinations of nearly-degenerate singlet-dimer coverings, and (iii) fluctuations of orphan spins, {\it etc\/}. Reflecting the spatially random character of the state, there would be no characteristic energy scale in these excitations, as postulated in a phenomenological theory of Ref. \cite{AndHalVar}. Such an analogy yields the $T$-linear low-$T$ specific heat for the random-singlet state \cite{Uematsu2}.

 The next question to be addressed might be whether the QSL state is ever possible in 3D, whatever its physical origin. Generally, 3D magnets tend to be subject to less fluctuations and to stronger ordering tendency than in 2D even on frustrated lattices. In this connection, the pyrochlore lattice, known to be a highly frustrated 3D lattice, might provide a promising stage.

 An intensively studied example might be Tb$_2$Ti$_2$O$_7$ possessing a modest amount of Ising-like $\langle 111\rangle$ anisotropy \cite{Gardner,Fennell,Lhotel,Fritsch,Takatsu}. This material has widely been regarded as a quantum analog of the classical spin ice, {\it i.e.\/}, quantum spin ice \cite{Gardner,Fennell,Takatsu}. Meanwhile, it turns out that even a minute amount of off-stoichiometry induces a sharp specific-heat peak signaling a thermodynamic phase transition \cite{Takatsu}. Furthermore, the onset of the glass-like spin freezing was observed in samples both with and without a sharp specific-heat peak \cite{Lhotel,Fritsch}. In any case, the underlying physics of Tb$_2$Ti$_2$O$_7$ would be that of the anisotropic spin ice.

 One may then ask what is the situation in more isotropic Heisenberg-like pyrochlore magnets. Recently, an interesting Heisenberg-like pyrochlore magnet exhibiting the QSL behavior down to the low temperature of 0.5K was reported in the mixed-anion antiferromagnet Lu$_2$Mo$_2$O$_5$N$_2$ \cite{Pattfield}, where the magnetism is borne by $s=1/2$ spins of Mo$^{5+}$ 4$d^1$, in contrast to its precursor oxide Lu$_2$Mo$_2$O$_7$ where the magnetism is borne by $s=1$ spins of Mo$^{4+}$ 4$d^2$. The oxynitride Lu$_2$Mo$_2$O$_5$N$_2$ was observed to exhibit gapless QSL behaviors characterized by the $T$-linear specific heat and the broad dynamical structure factor, in contrast to the oxide Lu$_2$Mo$_2$O$_7$ which exhibits a spin-glass (SG) freezing and the low-$T$ specific heat proportional to $T^2$. An interesting observation is that the oxynitride inevitably contains a significant amount of quenched disorder derived from the random occupation of O$^{2-}$/N$^{3-}$ anions. Hence, somewhat counter-intuitively, the introduction of randomness into the nominally disorder-free SG state apparently induces the QSL state \cite{Pattfield}. Lu$_2$Mo$_2$O$_5$N$_2$ was recently investigated theoretically by means of the density functional theory and the pseudofermion functional renormalization group method, suggesting the importance of the third-neighbor interaction $J_3$ in stabilizing the QSL state, while the analysis was basically that of the homogeneous system and the effect of O/N randomness was not considered explicitly \cite{Iqbal}.

 Under such circumstances, 
we wish to investigate in the present Letter the role of randomness in the quantum Heisenberg antiferromagnet on the 3D pyrochlore lattice. We consider the bond-random $s=1/2$ isotropic Heisenberg model on the pyrochlore lattice with the AF NN interaction. The Hamiltonian is given by
\begin{align}
\mathcal{H}=J\sum_{\Braket{i,j}}j_{ij}{\bm S}_i\cdot{\bm S}_j,
\label{eq:hamiltonian}
\end{align}
where ${\bm S}_i=(S_i^x, S_i^y, S_i^z)$ is an $s=1/2$ spin operator at the $i$-th site, and the sum $\Braket{i,j}$ is taken over all NN pairs on the lattice, while $j_{ij}\ge 0$ is the random variable obeying the bond-independent uniform distribution between $[1-\Delta, 1+\Delta]$ with $0\leq \Delta\leq 1$. We consider the NN interaction only just for simplicity, and put $J=1$ as the energy unit. The parameter $\Delta$ represents the extent of the randomness: $\Delta=0$ corresponds to the regular case and $\Delta=1$ to the maximally random case when the interaction is to be kept AF. 

 The corresponding regular model with $\Delta=0$ has been studied extensively by various numerical methods. There is a consensus that the system remains disordered even at $T=0$ without any spin LRO, while the nature of the nonmagnetic ground state still remains unclear \cite{Chandra}.
Whether the ground state is gapped or gapless also remains controversial, though the bulk of the numerical calculations seem to suggest a nonzero spin gap. By contrast, there has been no systematic numerical study of the corresponding random model with $\Delta>0$.

 We then study the ground-state properties of the model by means of the exact diagonalization (ED) Lanczos method. We treat finite-size clusters with the total number of spins $N$ up to $N\leq36$ ($N$ is taken to be a multiple of 4 with $8\leq N\leq 36$), periodic boundary conditions applied in all directions. The clusters of $N=16$ and $N=32$ possess the cubic symmetry of the bulk pyrochlore lattice. The numbers of independent bond realizations (samples) $N_s$ used in the sample average are for the order parameter, the spin gap, and the static spin structure factor $N_s=100$ and 5 for  $N=8$--$32$ and 36, respectively, whereas  for the dynamical spin structure factor $N_s=100$ and 50 for $N=16$ and 32, respectively. Error bars are estimated from sample-to-sample fluctuations.  The finite-temperature properties are computed by the Hams--de Raedt method \cite{HamsRaedt,supple}. The computation is performed for the size $N=32$, where the averaging is made over 10 initial vectors and 10 independent bond realizations. Error bars of physical quantities are estimated from the scattering over both samples and initial states by using the bootstrap method \cite{Bootstrap}.

\begin{figure}
  \begin{center}
    \includegraphics[width=\hsize]{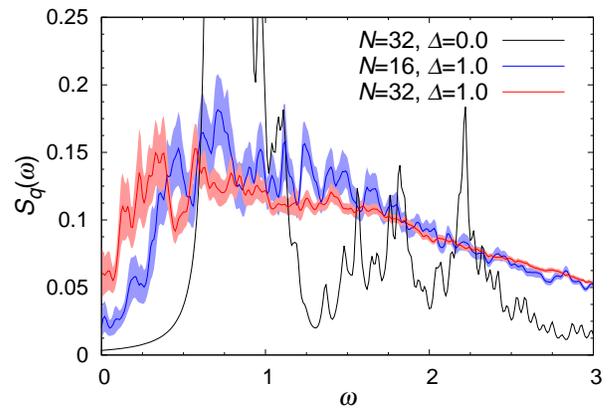}
   \caption{\label{fig:DSF} (Color online) The dynamical spin structure factor $S_{\bm q}(\omega)$ of the random model of $\Delta=1$ averaged over equivalent $(2,0,0)$, $(0,2,0)$, and $(0,0,2)$ points, as compared with the one of the regular model of $\Delta=0$. The lattice size is $N=16$ and 32. Error bars are represented by the width of the data curves.}
  \end{center}
\end{figure}
%


 We first investigate the existence or nonexistence of the magnetic LRO by computing the spin freezing parameter $\bar q$ defined by
\begin{align}
\bar{q}^2= \frac{1}{N^2}
\sum_{i,j}\left[\Braket{{\bm S}_i\cdot{\bm S}_j}^2 \right]_J,
\label{eq:q_bar}
\end{align}
where $[\cdots ]_J$ denotes the average over the disorder or samples \cite{SGreview}. This quantity can detect the static spin order of any type, even including the random one like the SG order.  The computed $\bar q$ is plotted versus $N^{-1/3}$ in Fig. \ref{fig:order}(a) for various values of randomness $\Delta$, in which the spin-wave form $\bar q_N\approx \bar q_\infty + cN^{-1/3}$ is borne in mind. As can be seen from the figure, $\bar q$ is extrapolated to zero, indicating the absence of spin LRO for any $\Delta$, even including the SG one. 

 In Fig. \ref{fig:order}(b), we show the size dependence of the spin-gap energy $\Delta E$. For smaller $\Delta < \Delta_c\simeq 0.6$, $\Delta E$ tends to be extrapolated to a nonzero value suggestive of a gapped behavior, whereas, for larger $\Delta> \Delta_c$, it is extrapolated to zero within the error bar suggestive of a gapless behavior. The changeover observed between the gapped and gapless behaviors suggests the occurrence of a randomness-induced phase transition between the two distinct types of nonmagnetic states. The randomness-induced gapless nonmagnetic phase stabilized at $\Delta > \Delta_c$ is likely to be the random-singlet state as identified in Refs. \cite{Watanabe,Kawamura,Shimokawa,Uematsu,Uematsu2}.
Although the precise value of $\Delta_c$ remains uncertain, we tentatively quote $\Delta_c=0.6^{+0.1}_{-0.3}$.

 Such a transition can also be detected via the quantity $R$. All the samples we have studied possess either singlet or triplet ground states, and $R$ is the ratio of the number of samples with triplet ground states expected to be related to the fraction of orphan spins accompanied by the divergent-like low-temperature susceptibility.

 As can be seen from Fig. \ref{fig:order}(c), $R$ vanishes for $\Delta < \Delta_c\sim0.6$, while it grows taking nonzero values beyond $\Delta_c$. This observation is consistent with a gapped-gapless transition observed in Fig. \ref{fig:order}(b), which strengthens our conclusion of a phase transition occurring between the randomness-irrelevant gapped QSL state and the randomness-relevant gapless QSL state. 
 
 In order to probe the properties of magnetic excitations, we also compute the dynamical structure factor $S_{\bm q}(\omega)$ given by \cite{Gagliano,Shimokawa}
\begin{align}
S_{\bm q}(\omega)&=-\lim_{\eta\to 0}\left[\frac{1}{\pi}{\rm Im}\Braket{
    (S_{\bm q}^z)^\dag\frac{1}{\omega + E_0 + i\eta -\mathcal{H}}
    S_{\bm q}^z}\right]_J ,
\label{eq:DSF}
\end{align}
where $E_0$ is the ground-state energy, and $\eta$ is a phenomenological damping factor taking a sufficiently small positive value. We employ the continued fraction method to compute $S_{\bm q}(\omega)$ \cite{Gagliano}, putting $\eta=0.02$.
 The $\omega$-dependence of the computed $S_{\bm q}(\omega)$ in the random-singlet state is shown for the case of the maximal randomness of $\Delta=1$ in Fig. \ref{fig:DSF} at the (2,0,0) $\bm{q}$-point (given in units of $2\pi/a$ where $a$ is the linear size of the cubic unit cell), at which the corresponding static spin structure factor $S_{\bm q}$ has relatively high intensity, in comparison with the corresponding data for the regular model of $\Delta=0$. The information of $S_{\bm q}$ exhibiting broad features in the ${\bm q}$-space is given in Fig. S1 of Supplemental Material \cite{supple}. As can be seen from Fig. \ref{fig:DSF}, the computed $S_{\bm q}(\omega)$ exhibits broad features in $\omega$ without any clear peak, accompanied by a long tail extending to larger $\omega$, indicating the absence of characteristic energy scale. We note that essentially similar $\omega$-dependence is observed for other ${\bm q}$-points as well. Such a feature is also common to the random-singlet states in 2D \cite{Kawamura,Shimokawa,Uematsu,Uematsu2}. In the small-$\omega$ region, $S_{\bm q}(\omega)$ diminishes somewhat toward $\omega \rightarrow 0$, though it still stays gapless as can be confirmed from its system-size dependence, in sharp contrast to the gapped behavior observed for $\Delta=0$.

\begin{figure}
  \begin{center}
    \includegraphics[width=\hsize]{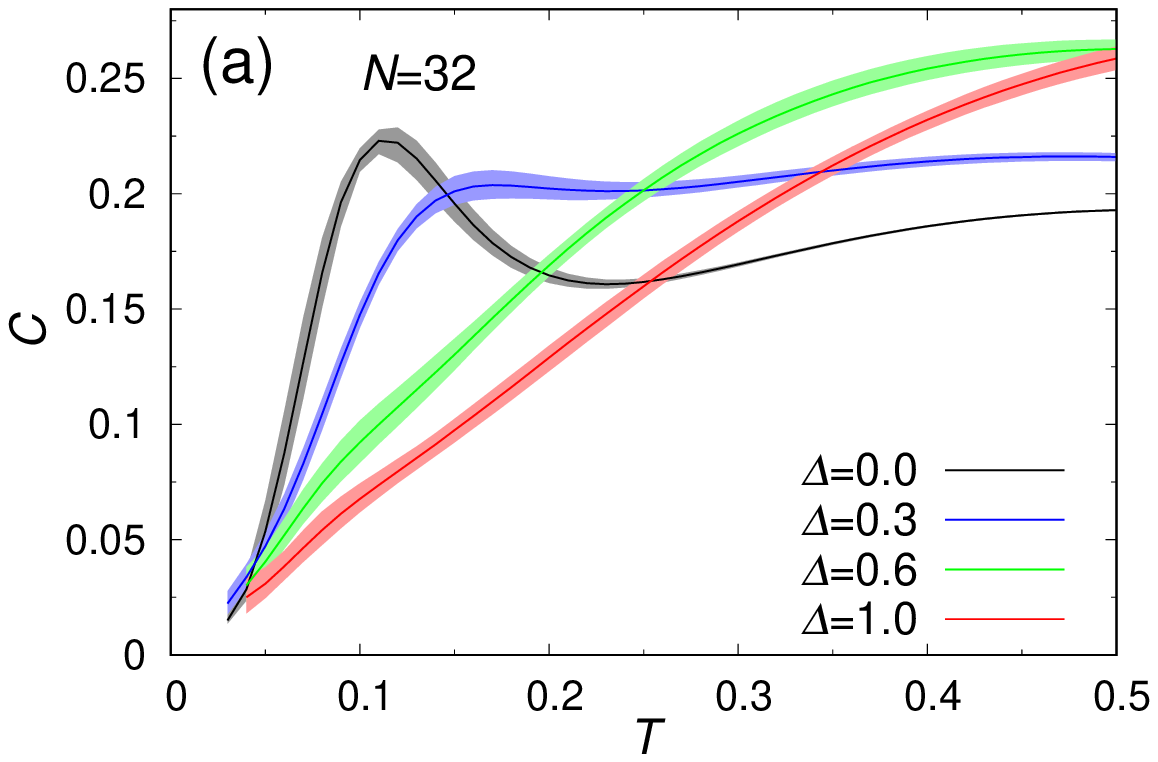}
    \includegraphics[width=\hsize]{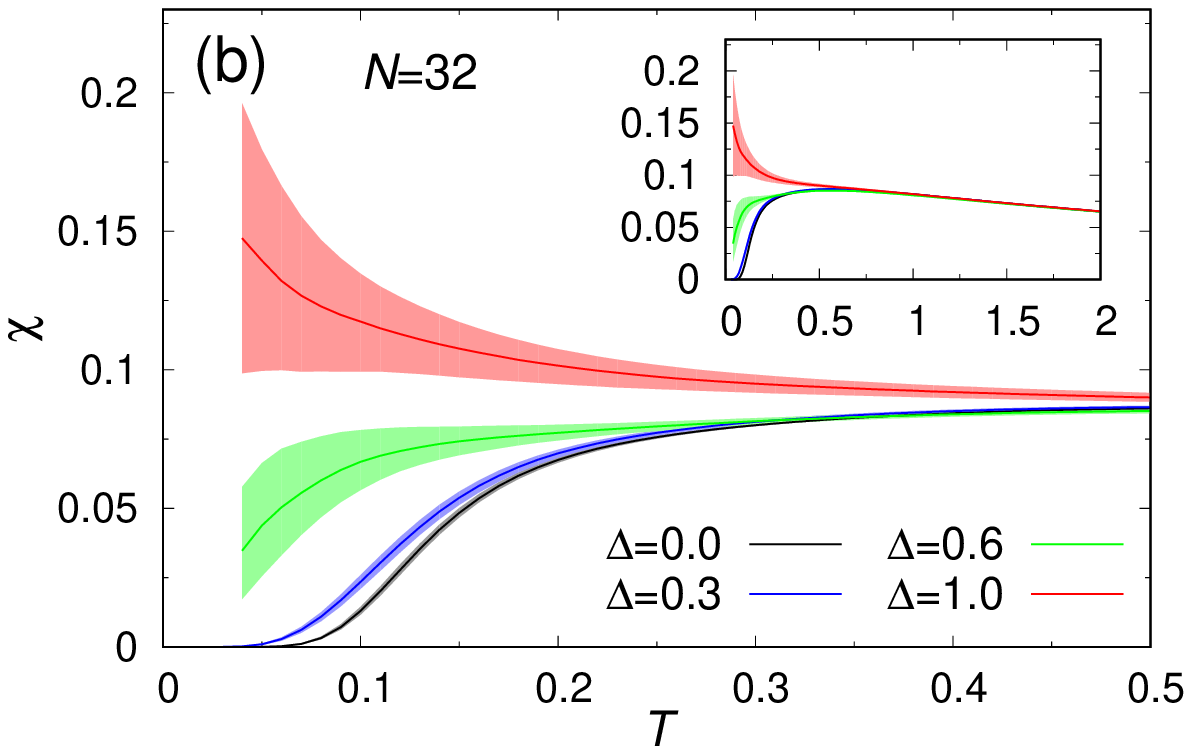}
    \caption{\label{fig:finiT-random} (Color online) The temperature dependence of (a) the specific heat per spin $C$, and of (b) the uniform susceptibility per spin $\chi$ (wider temperatue range in the inset), for various values of $\Delta$.}
  \end{center}
\end{figure}

We next move to the finite-$T$ properties. In Fig. \ref{fig:finiT-random}, we show the temperature dependence of (a) the specific heat per spin, and of (b) the susceptibility per spin. As can be seen from Fig. \ref{fig:finiT-random}(a), while the specific heat exhibits a double-peak structure for vanishing and weaker randomness $\Delta$, such a structure is gone in the random-singlet state at $\Delta \gtrsim 0.6$. We note that the double-peak structure generally means bunch of low-energy excitations with two distinct energy scales, which, however, is not the case in the random-singlet state. For $\Delta\gtrsim0.6$, as generically observed in the random-singlet states in 2D \cite{Watanabe,Kawamura,Uematsu,Uematsu2}, it exhibits a $T$-linear behavior at lower temperatures, $C\simeq\gamma T$, probably originated from the type of low-energy excitations postulated in \cite {Uematsu2,AndHalVar}. Such a change of behavior is also consistent with the occurrence of a phase transition within the nonmagnetic state at $\Delta=\Delta_c$ argued above. As can be seen from Fig. \ref{fig:finiT-random}(b), the susceptibility exhibits a nearly $T$-independent behavior insensitive to $\Delta$ for $0.2\lesssim T\lesssim 2$, while in the region of the random-singlet state of $\Delta \geq \Delta_c$ exhibits a gapless behavior with a Curie-like tail at still lower $T$, suggesting the existence of orphan spins \cite{Watanabe,Kawamura,Uematsu,Uematsu2}.

 One sees from our present results that the properties of the randomness-induced  QSL state, the random-singlet state, of the 3D pyrochlore Heisenberg model is rather similar to those of the 2D models \cite{Watanabe,Kawamura,Shimokawa,Uematsu,Uematsu2}. This observation may suggest that the nature of the random-singlet state is insensitive not only to the details of the lattice structure and the origin of frustration \cite{Watanabe,Kawamura,Shimokawa,Uematsu,Uematsu2}, but even to whether the spatial dimensionality is either two or three. In other words, the random-singlet state seems to be a universal state of magnets. Of course, system size studied here are very small, and care needs to be taken in reaching a definitive conclusion.

While the random-singlet features are expected to be most eminent for stronger randomness, being clearly visible even for smaller systems and at higher temperatures, they might be realized also for somewhat weaker randomness,  manifesting themselves only at longer length scales and at lower temperatures not directly accessible by the present ED calculation (recall the large uncertainty in our estimate of $\Delta_c$), even with possible experimental relevance expected \cite{Kimchi,Kimchi2}. Such a random-singlet-like state for weaker randomness is an adiabatic continuation of the one for stronger randomness so long as it remains gapless, and is essentially the same state as that for stronger randomness.

 We now wish to discuss experimental implications.  Our results are compared favorably with the experimental results on Lu$_2$Mo$_2$O$_5$N$_2$ \cite{Pattfield}, at least qualitatively, {\it i.e.\/}, the $T$-linear specific heat, the gapless susceptibility accompanied by an intrinsic Curie-like tail, and the broad spectrum of $S_{\bm q}(\omega)$. Quantitatively, however, some deviation remains. If we try to estimate the coefficient of the $T$-linear specific heat with the $J$-value deduced from the experimental Curie-Weiss temperature, we get $\gamma \sim 80$ mJ/molK$^2$ from our result, which deviates considerably from the experimental value $\gamma\sim10$ mJ/molK$^2$ \cite{Pattfield}. One possible cause might be our oversimplified assumption of the randomness. The other possibility is that further neighbor interactions \cite{Iqbal}, especially $J_3$, might affect the underlying energetics at the quantitative level.  Unfortunately, the system size accessible by the ED method is too small to take account of such $J_3$-effect in a meaningful manner. Anyway, the random-singlet state is stabilized as long as the system possesses a certain amount of randomness, frustration and quantum fluctuations, staying quite robust at the qualitative level against other details of the system.


 In summary, we studied both the $T=0$ and $T>0$ properties of the bond-random $s=1/2$ NN Heisenberg model on the 3D pyrochlore lattice, and found that the randomness-induced gapless QSL state, the random-singlet state, is stabilized if the strength of the randomness exceeds a critical value. Its properties turn out to be rather similar to those of the frustrated random Heisenberg magnets in 2D, highlighting the possible universal character of the random-singlet state. Further studies are desirable to fully substantiate such a picture.

\begin{acknowledgements}
 The authors wish to thank Y. Iqbal, H.O. Jeschke and K. Aoyama for valuable discussion. This study was supported by JSPS KAKENHI Grant Number JP25247064. Our code was based on TITPACK Ver.2 coded by H. Nishimori. We are thankful to ISSP, the University of Tokyo, and to YITP, Kyoto University, for providing us with CPU time.
\end{acknowledgements}


\end{document}



\title{Supplemental Material: Randomness-induced quantum spin liquid behavior in the $s=$1/2 bond-random Heisenberg antiferromagnet on the pyrochlore lattice}


\author{Kazuki Uematsu}
\author{Hikaru Kawamura}

\maketitle

\section*{Spin freezing parameter}
The spin freezing parameter $\bar q$ is defined by Eq. (2) in the main text, {\it i.e.\/},
\begin{align}
  \bar{q}^2=\frac{1}{N^2}\sum_{i,j}\left[\braket{\bm{S}_i\cdot\bm{S}_j}^2\right]_J.
  \label{eq:qbar}
\end{align}
This quantity takes a nonzero value in the spin ordered state, even including the spatially random one such as spin glasses, and vanishes in the paramagnetic state. Thus, it serves as the order parameter of the spin ordering or freezing. Below, we shall explain how this is the case.

 Usually, the existence of the magnetic long-range order can be detected via the two-point spin correlation function $g({\bm r}_{ij})=\left[ \langle {\bf S}_i\cdot {\bm S}_j\rangle \right]_J$ by looking at its long-distance behavior, {\it i.e.\/}, whether $m^2\equiv \lim_{|r_{ij}|\rightarrow \infty} g({\bm r}_{ij})$ is zero or nonzero. In fully random systems, this quantity $m^2$ trivially vanishes after the configuration or sample average due to the sign cancellation of $\langle {\bf S}_i\cdot {\bm S}_j\rangle$, and does not work as an appropriate order parameter. This problem can be avoided simply by squaring $\langle {\bf S}_i\cdot {\bm S}_j\rangle$  before the configurational average to make it non-negative. Then, an appropriate two-point correlation function might be
\begin{align}
  g^{(2)}({\bm r}_{ij})=\left[ \braket{\bm{S}_i\cdot\bm{S}_j}^2 \right]_J.
 \label{eq:qbar}
\end{align}
This correlation function decays exponentially with a finite correlation length in the paramagnetic state, but tends to a nonzero value in the long-distance limit in the magnetically ordered state, even including the spatially random ordered state. Then, the long-distance limit of $g^{(2)}({\bm r}_{ij})$,
\begin{equation}
  \bar q_{\infty}^2 = \lim_{|r_{ij}|\rightarrow \infty} g^{(2)}({\bm r}_{ij})
\end{equation}
yields an appropriate spin freezing parameter in the thermodynamic limit.

Equivalently, if one defines the finite-size spin freezing parameter by Eq. (\ref{eq:qbar}) above, it reduces in the thermodynamic limit $N\rightarrow \infty$ to the spin freezing parameter $\bar q_\infty$ defined above.

 This spin freezing parameter can also be related to the so-called Edward-Anderson order parameter \cite{EdwardsAnderson} or the spin overlap well-known in spin-glass studies: see Ref. \cite{SGreview} for further details.

\section*{The Hams--de Raedt method}
The Hams-de Raedt method obtains the finite-temperature properties of the model with its Hamiltonian $\mathcal{H}$ based on the Taylor expansion of the Boltzmann factor $\exp(-\beta\mathcal{H})$ \cite{Hams-deRaedt,cTPQ}. The thermal average of the physical quantity $A$ can be represented as
\begin{align}
\braket{A}&=
\frac{ {\rm Tr}[\exp(-\beta\mathcal{H}/2) A \exp(-\beta\mathcal{H}/2)]} {{\rm Tr}[\exp(-\beta\mathcal{H}/2) \exp(-\beta\mathcal{H}/2)]} \nonumber
\\
&=\frac{ \overline{\Braket{0|\exp(-\beta\mathcal{H}/2) A \exp(-\beta\mathcal{H}/2)|0}}} {\overline{\Braket{0|\exp(-\beta\mathcal{H}/2) \exp(-\beta\mathcal{H}/2)|0}}} \nonumber
\\
&=\frac{ \overline{\Braket{0|(\sum_k (-\beta\mathcal{H}/2)^k/k!) A (\sum_{k'} (-\beta\mathcal{H}/2)^{k'}/k'!)|0}}} {\overline{\Braket{0|(\sum_k (-\beta\mathcal{H}/2)^k/k!) (\sum_{k'} (-\beta\mathcal{H}/2)^{k'}/k'!)|0}}},
\label{eq:HdR1}
\end{align}
where $k=0,1,2,\cdots$ is a non-negative integer and $\overline{\cdots}$ is the average over the random initial vectors $\ket{0}$. The term $(-\beta\mathcal{H}/2)^k/k!$  becomes very large at low temperatures. To control this divergent behavior, an appropriate normalization factor is introduced. Let $E_{min}$ the minimum (ground-state) energy eigenvalue,  $E_{max}$ the maximum energy eigenvalue, and $E_w$ their difference $E_w=E_{max}-E_{min}$. Then, with the use of the convergence factor $h=(E_{max}-\mathcal{H})/E_w$, Eq. (\ref{eq:HdR1}) can be rewritten as
\begin{align}
\braket{A}=\frac{\overline{\sum_{k,k'} \Braket{k|A|k'}}}{\overline{\sum_{k,k'} \Braket{k|k'}}},
\label{eq:HdR2}
\end{align}
where the state $\ket{k}$ is given by
\begin{align}
\ket{k}=\exp(-\beta E_w/2) (\beta E_w/2)^k/k!\cdot h^k\ket{0}.
\label{eq:HdR3}
\end{align}
We compute the physical quantities based on Eqs. (\ref{eq:HdR2}) and (\ref{eq:HdR3}).

\begin{figure}[t]
  \begin{tabular}{c}
    \begin{minipage}{0.5\hsize}
      \includegraphics[width=\hsize]{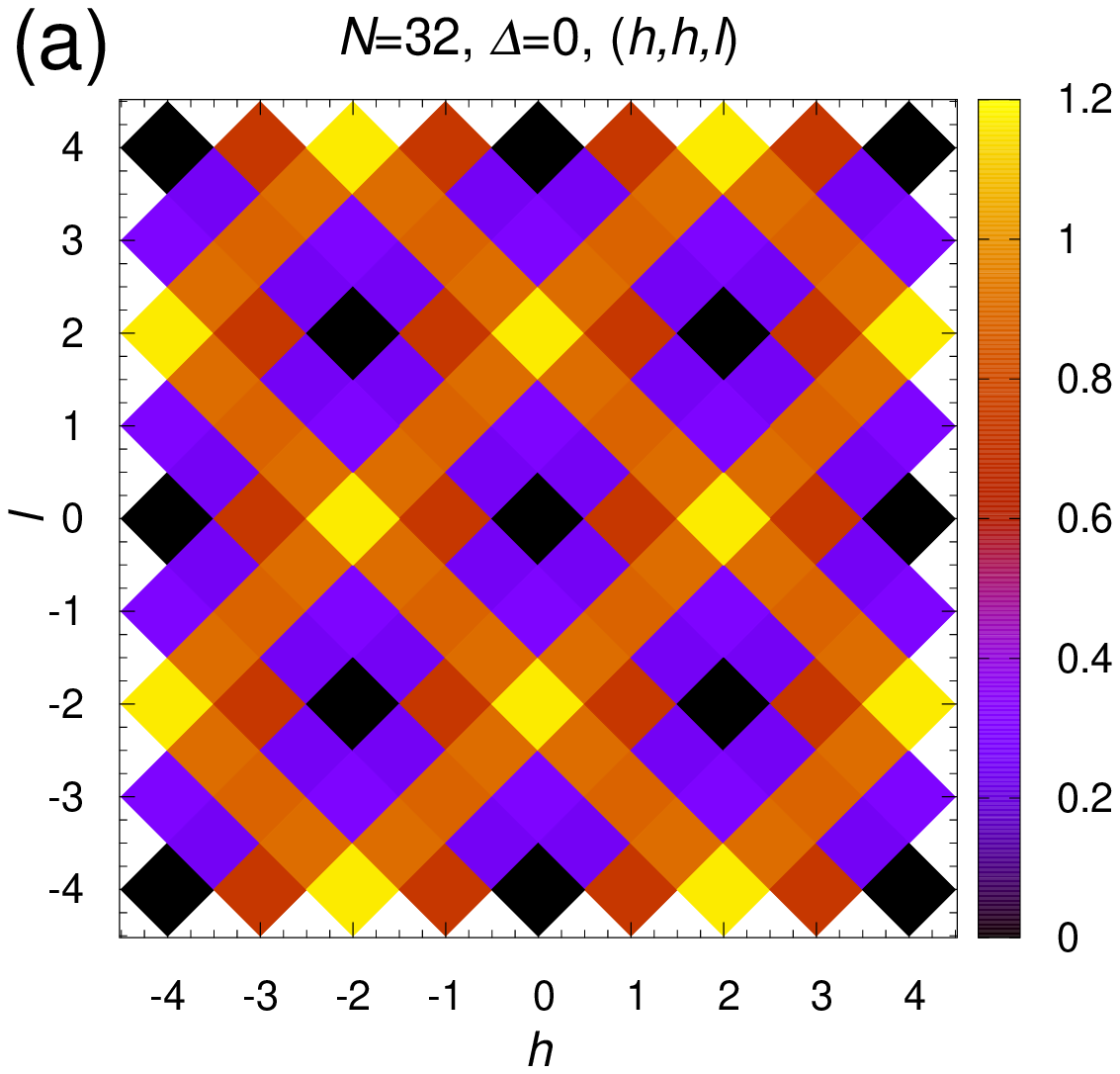}
    \end{minipage}
    \begin{minipage}{0.5\hsize}
      \includegraphics[width=\hsize]{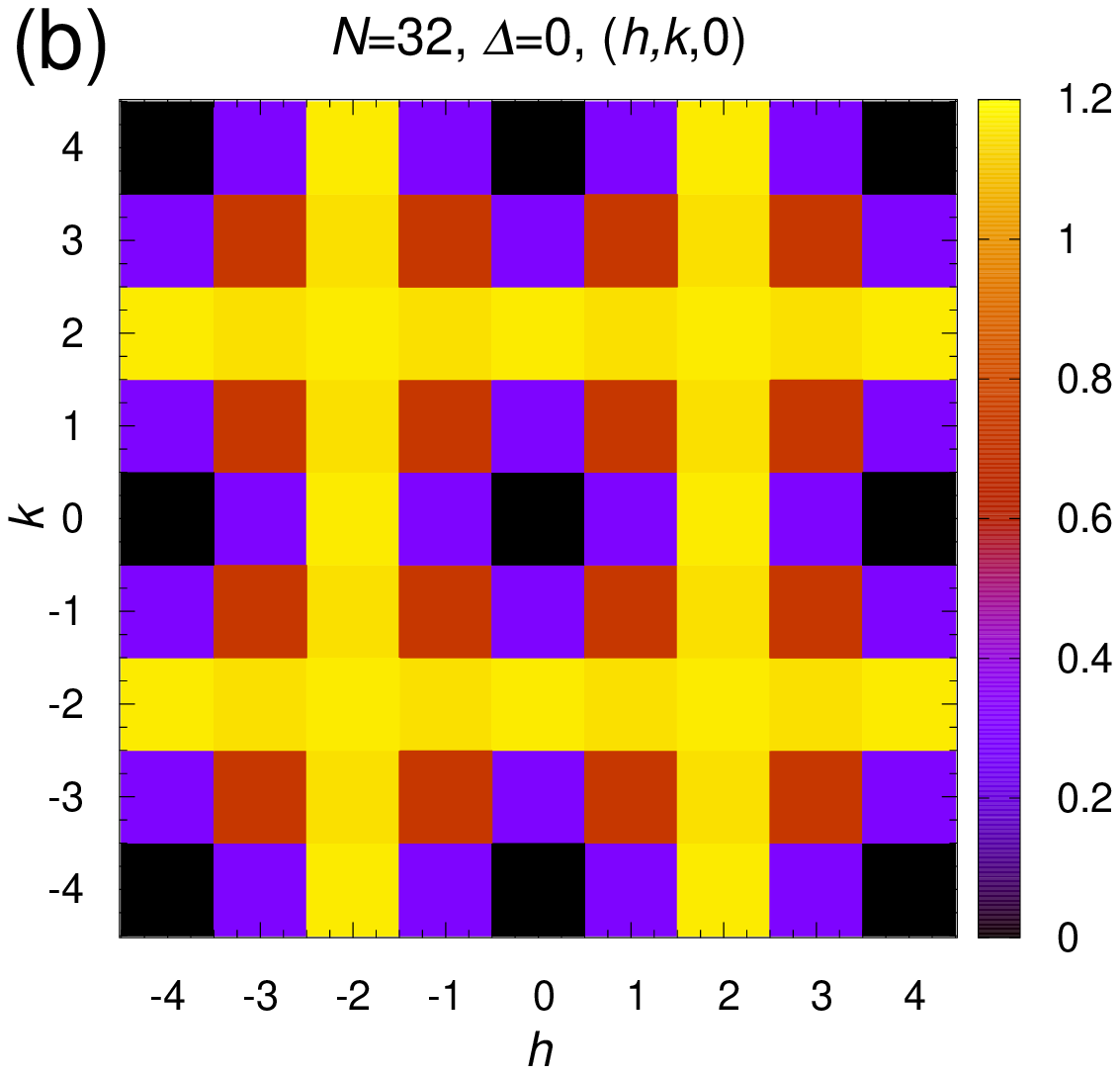}
    \end{minipage}\\
    \begin{minipage}{0.5\hsize}
      \includegraphics[width=\hsize]{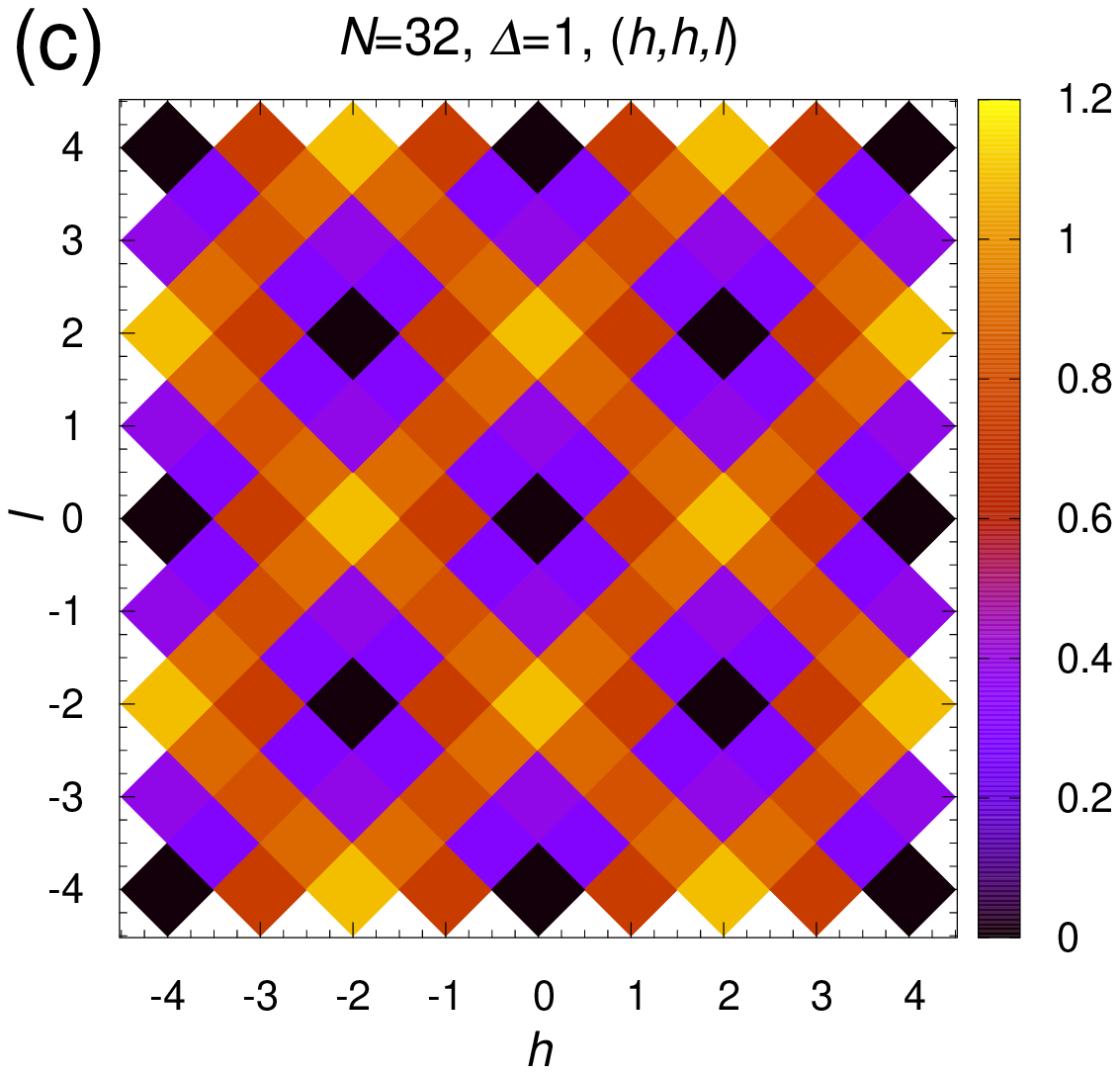}
    \end{minipage}
    \begin{minipage}{0.5\hsize}
      \includegraphics[width=\hsize]{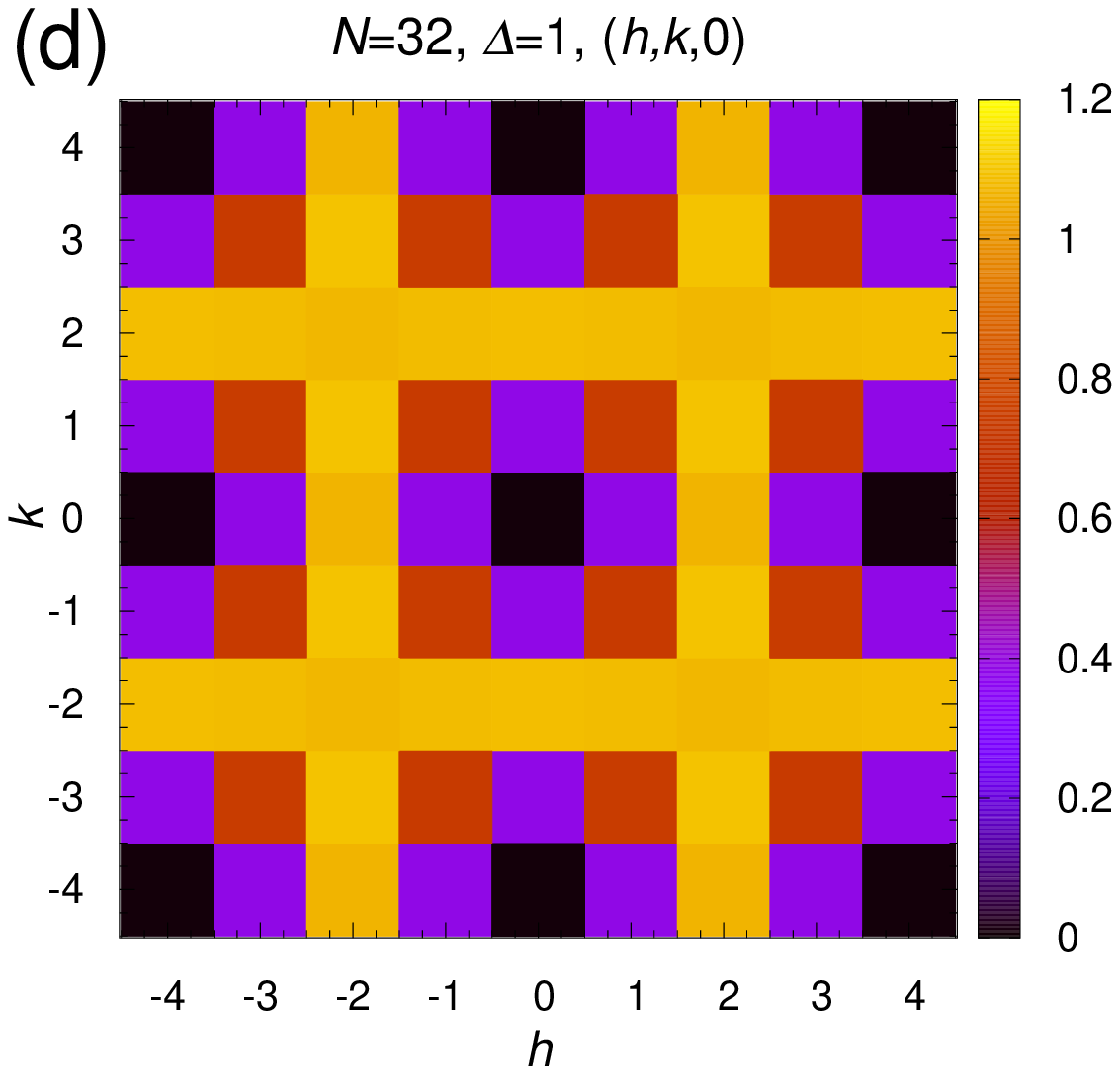}
    \end{minipage}
  \end{tabular}
  \caption{\label{fig:SSF} (Color online) 
    Upper row: Static spin structure factor $S_q$ for the regular case
    of $\Delta=0$ in the (a) $(h,h,l)$ and (b) $(h,k,0)$ plane.
    Lower row: Static spin structure factor $S_q$ for the maximally random case
    of $\Delta=1$ in the (c) $(h,h,l)$ and (d) $(h,k,0)$ plane.
  }
\end{figure}

\section*{Static spin structure factor}
The computed ground-state spin structure factor $S_{\bm q}$ is defined by
\begin{align}
S_{\bm q}&=\frac{1}{N}\left[ \Braket{ |{\bm S}_{\bm q}|^2}\right]_J
\nonumber \\
&=\frac{1}{N} \left[ \sum_{i,j} \Braket{{\bm S}_i\cdot{\bm S}_j}
  \cos{\left(\frac{2\pi}{a}{\bm q}\cdot\left({\bm r}_i-{\bm r}_j\right)\right)}
  \right]_J,
\label{eq:SSF}
\end{align}
where ${\bm S}_{\bm q}=\sum_j {\bm S}_j e^{i\frac{2\pi}{a}{\bm q}\cdot{\bm r}_j}$ is the Fourier transform of the spin operator, ${\bm r}_j$ is the position vector at the site $j$, ${\bm q}$ is the wave vector, $a$ is the linear size of the cubic unit cell of the pyrochlore lattice, while $\langle \cdots \rangle$ and $[\cdots]_J$ represent the ground-state expectation value (or the thermal average at finite temperatures) and the configurational average over $J_{ij}$ realizations.

In Fig. \ref{fig:SSF}, the static spin structure factor for the regular case of $\Delta=0$ is shown in the upper row, while the one for the maximally random case of $\Delta=1$ is in the lower row.
The left side of Fig. \ref{fig:SSF} is the intensity plot in the $(h,h,l)$ plane, while the right side is in the $(h,k,0)$ plane.
As can be seen from these figures, there are no sharp peaks in any of Fig. \ref{fig:SSF}, indicating the absence of the magnetic long-range order.
The difference between the upper and the lower rows turns out to be relatively minor so that the introduction of randomness seems to give minor effects on the static spin structure factor.
This might partly be due to the finite-size effect, as the very coarse mesh size available here for $S_{\bm q}$ tends to mask the possible structure in the ${\bm q}$-space. In order to probe the possible difference between the regular and the random cases, much larger system sizes might be required. This is in contrast to the $\omega$-dependence of the dynamical structure factor $S_{\bm q}(\omega)$ where the finite-size effect tends to be more indirect and the difference between the regular and the random cases is much clearer, as shown in Fig. 2 of the main text.

 In any case, a certain nontrivial structure in the ${\bm q}$-space is discernible for the static $S_{\bm q}$ even in the random case, suggesting that the underlying spin structure possesses spatial correlations extending at least to several lattice spacings. This actually seems consistent with the picture of the random-singlet state where modest length-scale of objects are expected as its basic ingredients. Recall that, in the random-singlet state, some of the singlets are formed between distant neighbors and even the resonance between distinct singlet-dimer coverings are expected, as discussed in the main text.